\documentclass[reprint,superscriptaddress,floatfix,amsmath,amssymb,aps,prstab]{revtex4-1}

\usepackage{graphicx}
\usepackage{textcomp}
\usepackage{xcolor}
\usepackage[mathlines]{lineno} 
\definecolor{light-gray}{gray}{0.8}
\definecolor{prlblue}{rgb}{0.176, 0.152, 0.57}

\usepackage[colorlinks=true, linkcolor=black, citecolor=prlblue, filecolor=black, urlcolor=prlblue]{hyperref}

\begin{document}

\title{Matching small $\beta$ functions using centroid jitter and two beam position monitors}

\author{C.~A.~Lindstr{\o}m}
\email{carl.a.lindstroem@desy.de}
\author{R.~D'Arcy}
\author{M.~J.~Garland}
\affiliation{Deutsches Elektronen-Synchrotron DESY, Notkestra{\ss}e 85, 22607 Hamburg, Germany}
\author{P.~Gonzalez}
\affiliation{Deutsches Elektronen-Synchrotron DESY, Notkestra{\ss}e 85, 22607 Hamburg, Germany}
\affiliation{Universit{\"a}t Hamburg, Luruper Chaussee 149, 22761 Hamburg, Germany}
\author{B.~Schmidt}
\affiliation{Deutsches Elektronen-Synchrotron DESY, Notkestra{\ss}e 85, 22607 Hamburg, Germany}
\author{S.~Schr{\"o}der}
\affiliation{Deutsches Elektronen-Synchrotron DESY, Notkestra{\ss}e 85, 22607 Hamburg, Germany}
\affiliation{Universit{\"a}t Hamburg, Luruper Chaussee 149, 22761 Hamburg, Germany}
\author{S.~Wesch}
\author{J.~Osterhoff}
\affiliation{Deutsches Elektronen-Synchrotron DESY, Notkestra{\ss}e 85, 22607 Hamburg, Germany}



\begin{abstract}
Matching to small beta functions is required to preserve emittance in plasma accelerators. The plasma wake provides strong focusing fields, which typically require beta functions on the mm-scale, comparable to those found in the final focusing of a linear collider. Such beams can be time-consuming to experimentally produce and diagnose. We present a simple, fast, and noninvasive method to measure Twiss parameters in a linac using two beam position monitors only, relying on the similarity of the beam phase space and the jitter phase space. By benchmarking against conventional quadrupole scans, the viability of this technique was experimentally demonstrated at the FLASHForward plasma-accelerator facility.
\end{abstract}

\maketitle

\section{Introduction}
Plasma-wakefield accelerators \cite{TajimaDawsonPRL1979,ChenPRL1985,RuthPA1985} can provide accelerating gradients in the GV/m-range \cite{BlumenfeldNature2007,LitosNature2014}, promising smaller and cheaper accelerators \cite{JoshiPT2003,LeemansPT2009}. Reaching high energies, needed for X-ray free-electron lasers \cite{MadeyJAP1971,CouprieJPB2014} and linear colliders \cite{RosenzweigNIMA1998,SchroederPhysRevSTAB2010,AdliSnowmass2013,LindstromThesis2019} in particular, will require multiple accelerator stages \cite{SteinkeNature2016,LindstromNIMA2016} and hence some form of external beam injection into the plasma wake. 

Since the focusing field from an exposed ion column in a plasma accelerator is typically very strong, beams must be tightly focused for the beam size not to oscillate, as this would lead to significant and unacceptable emittance growth \cite{MehrlingPRAB2012}. In terms of Twiss or Courant-Snyder parameters \cite{CourantSnyderAoP1957}, the beta function needs to be matched to
\begin{equation}
    \label{eq:MatchedBeta}
    \beta_m = \sqrt{\frac{2 E \epsilon_0}{n e^2}}
\end{equation}
where $E$ is the beam energy, $n$ is the plasma density, $\epsilon_0$ is the vacuum permittivity, and $e$ is the electron charge. Injecting a GeV-level beam into a typical plasma accelerator requires beta functions on the mm-scale. While plasma density ramps \cite{MarshPAC2005,DornmairPRAB2015,ZhaoPRAB2020} can relax the matching condition by increasing $\beta_m$ at the entry and exit of the accelerator stage, it will nevertheless be challenging and time-consuming to experimentally produce and diagnose the required tightly-focused beams.

Conventional beam-focus diagnostics include wire scanners and high-resolution screens around the focal point, or downstream quadrupoles that point-to-point image the beam onto a screen---all of which require nontrivial experimental setups and careful data analysis. This can be inconvenient when matching beams into a plasma accelerator---a slow multi-parameter optimization process where fast feedback will be crucial.

In this paper, we present an alternative method for simple, fast, and noninvasive measurement of small beta functions by using two beam position monitors (BPMs) to measure the centroid jitter. The technique is based on the observation that the phase space of the jitter often has similar Twiss parameters to that of the beam, and can therefore be used as a proxy. While the method is approximate in nature, it allows online monitoring and iterative adjustment of the waist location and beta function. This technique was successfully implemented and experimentally demonstrated at the FLASHForward \cite{AschikhinNIMA2016,DArcyRSTA2019} plasma-accelerator facility at DESY.

\section{Beam and jitter phase spaces}
\label{sec:PhaseSpaces}

\begin{figure*}[t]
	\centering\includegraphics[width=0.98\textwidth]{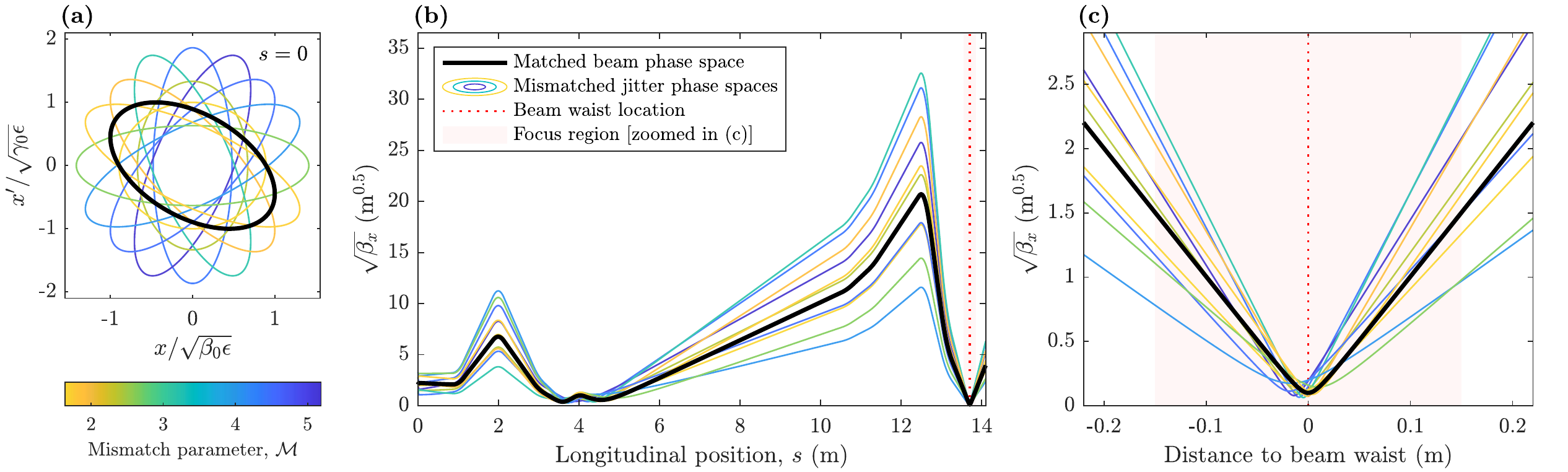}
	\caption{Evolution of a matched beam phase space and various possible mismatched jitter phase spaces through a strong-focusing lattice. Starting out moderately mismatched (a), the jitter beta functions evolve (b) and appear to diverge from that of the matched beam beta function. Nevertheless, in the focus region (c) the mismatched jitter phase spaces are all focused to a similar waist beta function and waist location as the beam.}
    \label{fig:Fig1}
\end{figure*}

The phase space of a beam consists of its particle distribution in $x$--$x'$ space (in one transverse plane). Similarly, the phase space of the beam centroid jitter---the \textit{jitter phase space}---is the distribution of beam centroid offsets in $x$--$x'$ space when integrated over a large number of shots. Therefore, the jitter has its own Twiss parameters and emittance.

The central assumption underpinning this technique is that the Twiss parameters of the jitter are similar to those of the beam. A significant consequence of this connection is that it is possible to simply and noninvasively measure the phase space of the jitter using BPMs, which then acts as an approximate measurement of Twiss parameters of the beam. It should be noted that this technique is generally not suitable for measuring the beam emittance, but this is also not required for matching (see Eq.~\ref{eq:MatchedBeta}).

While the similarity of the beam and the jitter is not guaranteed, it is motivated by both experimental observation and theoretical considerations. Linear accelerators usually have long FODO-like lattices with beta functions on the 1--10~m scale. This means that magnets and accelerating cavities---sources of jitter---are typically distributed across a range of phase advances. As a consequence, the jitter-phase-space ellipse gradually expands while it rotates to acquire a similar shape to the beam-phase-space ellipse. Conversely, if there were only a few dominant jitter sources---such as the gun or a single vibrating quadrupole---the jitter-phase-space ellipse would be disproportionately stretched in the $x'$-dimension at these phases. Similar beam and jitter-phase-space ellipses can therefore be expected in any well-commissioned machine where such dominant jitter sources have been removed. Even if the beam and jitter phase spaces are moderately mismatched, both will evolve and be focused similarly in a linear-optics lattice---also in the case of strong focusing, as demonstrated by the example in Fig.~\ref{fig:Fig1}. It should be noted that in the presence of strong focusing, the final-focusing quadrupoles can contribute disproportionately to the jitter phase space due to large beta functions---special care therefore needs to be taken to ensure their stability \cite{BalikIPAC2017,BettNIMA2018}.

The most interesting quantities in the context of matching are the location and beta function of the focus waist. How inaccurate should we expect the jitter-based measurement to be? Consider a lattice that focuses the beam to a small waist, where the beam size is demagnified by a factor $B$. Starting from matched Twiss parameters $\beta_0$ and $\alpha_0$, the resulting waist beta function would be $\beta_0/B^2$. The transfer matrix of such a lattice \cite{CourantSnyderAoP1957} can be expressed as
\begin{equation}
    R = \begin{bmatrix} \frac{\cos\psi+\alpha_0\sin\psi}{B} &  \frac{\beta_0}{B}\sin\psi \\  \frac{B}{\beta_0}\left(\alpha_0\cos\psi - \sin\psi\right) &  B \cos\psi \end{bmatrix},
\end{equation}
where the phase advance $\psi$ is a free parameter. Consider then a mismatched jitter with an initial betatron amplitude matrix
\begin{equation}
    \Sigma_0 = \begin{bmatrix} \beta &  -\alpha \\  -\alpha &  \gamma \end{bmatrix},
\end{equation}
where $\gamma = (1+\alpha^2)/\beta$ is the Twiss gamma function. The overall mismatch can be quantified by the \textit{mismatch parameter} \cite{SandsSLAC1991}
\begin{equation}
    \mathcal{M} = \frac{1}{2} \left(\tilde{\beta}_e + \tilde{\gamma}_e + \sqrt{(\tilde{\beta}_e + \tilde{\gamma}_e)^2 - 4} \right),
\end{equation}
where $\tilde{\beta}_e=\beta/\beta_0$, $\tilde{\alpha}_e=\alpha-\alpha_0\beta/\beta_0$ and $\tilde{\gamma}_e=(1+\tilde{\alpha}_e^2)/\tilde{\beta}_e$ quantify the normalized error of each Twiss parameter. The mismatch parameter $\mathcal{M}$ is invariant in a linear-optics lattice, whereas the individual Twiss errors are not.

We can transport the mismatched jitter to the beam waist location (i.e., the end of the lattice) using
\begin{equation}
    \Sigma = R \Sigma_0 R^T.
\end{equation}
The $\Sigma_{11}$ element corresponds to the jitter beta function at the beam waist location. However, the beam waist does not generally coincide with the jitter waist, and therefore $\Sigma_{11}$ does not correspond to the waist beta function of the jitter. Instead, assuming that the focus region consists only of a drift, the waist beta function equates to the inverse gamma function ($1/\Sigma_{22}$), which can be expressed as 
\begin{equation}
    \label{eq:MismatchedWaistBeta}
    \beta_w = \frac{\beta_0}{B^2} \left( \frac{1}{\tilde{\alpha}_e\sin2\psi + \tilde{\beta}_e\sin^2\psi + \tilde{\gamma}_e\cos^2\psi} \right).
\end{equation}
Similarly, the shift of the jitter waist location is given by the ratio of the alpha and the gamma function ($-\Sigma_{12}/\Sigma_{22}$), which is derived to be
\begin{equation}
    \label{eq:MismatchedWaistLocation}
    \delta s_w = \frac{\beta_0}{B^2} \left(  \frac{\tilde{\alpha}_e\cos2\psi+\frac{1}{2}(\tilde{\beta}_e-\tilde{\gamma}_e)\sin2\psi}{\tilde{\alpha}_e\sin2\psi + \tilde{\beta}_e\sin^2\psi + \tilde{\gamma}_e\cos^2\psi} \right).
\end{equation}
Although lengthy, it is easy to see that if the Twiss errors $\tilde{\alpha}_e$, $\tilde{\beta}_e$, and $\tilde{\gamma}_e$ are all of order one (i.e., moderately mismatched), the brackets in both Eqs.~\ref{eq:MismatchedWaistBeta} and \ref{eq:MismatchedWaistLocation} become numerical factors also of order one, regardless of the free parameter $\psi$. This means that the waist beta function of the mismatched jitter remains similar to the waist beta function $\beta_0/B^2$ of the matched beam. Moreover, it implies that the offset of the waist location is also approximately $\beta_0/B^2$---of the order of the waist beta function itself.

In a plasma accelerator, this mismatch leads to an emittance growth for beams of finite energy spread, as the phase-space ellipse of each energy slice rotates at a different rate. Fully decohered, the relative emittance growth saturates at \cite{MehrlingPRAB2012}
\begin{equation}
    \label{eq:MismatchEmitGrowth}
    \frac{\epsilon_{\textmd{sat}}}{\epsilon_0} = \frac{1}{2}\left(\mathcal{M} + \frac{1}{\mathcal{M}} \right),
\end{equation}
which also agrees with simulations. This implies that for a moderate mismatch ($\mathcal{M}$ of order one) the emittance growth is relatively small---e.g., a mismatch of $\mathcal{M}=2$ leads to an emittance growth of only 25\%. Using the jitter as a proxy is therefore appropriate for a quick first-pass matching to the plasma, before a final in-situ optimization using the beam.

\section{Two-BPM measurement method}

\begin{figure}[t]
	\centering\includegraphics[width=0.95\linewidth]{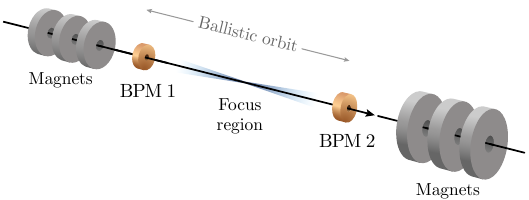}
	\caption{Basic experimental setup, with two BPMs surrounding the focus region, separated by a ballistic orbit. The measurement can also be generalized to nonballistic orbits, where there are magnets also between the BPMs.}
    \label{fig:Fig2}
\end{figure}

Having connected the phase space of the beam to that of the jitter, the problem has been reduced to measuring the jitter phase space. This can be done quickly and noninvasively with a multi-shot measurement using two BPMs---see Fig.~\ref{fig:Fig2} for a conceptual setup. Correlated offset data is required to measure the position and angle of each shot, which for a ballistic orbit (i.e., no magnets between the BPMs) is given by
\begin{equation}
    \label{eq:BPMangle}
    x' = \frac{x_2 - x_1}{\Delta s},
\end{equation}
where $x_1$ and $x_2$ are the upstream and downstream centroid offsets, respectively, and $\Delta s$ is the separation of the two BPMs. Both transverse planes can be measured simultaneously. As the number of shots increases, the jitter phase space will gradually build up, assuming that the optics remains unchanged. Whenever the optics \textit{does} change, the measurement must be restarted.

Given that no scan is performed, the data can be analyzed immediately from the start of the measurement, then re-analyzed with every additional shot, gradually increasing the precision. As the number of shots $N$ increases, the relative measurement error of Twiss parameters and jitter emittance will be approximately $1/\sqrt{N}$. Since the connection between the beam and the jitter phase space is only approximate, it will rarely be necessary to require better than about 10\% precision (corresponding to 50--500~shots). In a typical accelerator with a 1--10~Hz repetition rate, this allows quasi-online monitoring on a few-tens-of-seconds timescale. 

The measurement can also be generalized to nonballistic orbits (i.e., with magnets between the BPMs), as is relevant to for instance plasma accelerators with strong permanent quadrupoles close to the plasma entrance \cite{YaminIPAC2019}. In this case, the angle at the upstream BPM can be calculated using
\begin{equation}
    \label{eq:BPMgeneralAngle}
    x_1' = \frac{x_2 - M_{11} x_1}{M_{12}},
\end{equation}
where $M$ is the transfer matrix between the two BPMs. However, predicting the evolution of beta functions with mm-level accuracy then requires very accurate (per-mille-level) measurements of monitor locations, quadrupole locations, field strengths and beam energy---just like for a quadrupole scan. A ballistic measurement is comparatively simple, and hence always preferable if possible, as only an accurate measurement of BPM locations (and relative-offset calibrations) is required. For the remainder of this paper we will, therefore, assume that the BPMs are separated by only a drift space in order to facilitate ballistic measurements.

\section{Resolution limits}

The main limitation of this technique stems from the finite resolution of BPMs. In measuring the jitter-phase-space ellipse, the width of each angle-slice (i.e., the position jitter at the waist) must be well resolved, which limits how small a waist beta function can be measured.

\subsection{Analytic model}

To calculate this resolution limit, we consider the apparent covariance matrix of the jitter at the upstream BPM location
\begin{equation}
    \label{eq:CovarianceMatrix}
    \textmd{cov}(x,x') = \begin{bmatrix} \langle x^2 \rangle + \sigma^2 & \langle x x' \rangle - \frac{\sigma^2}{\Delta s} \\ \langle x x' \rangle - \frac{\sigma^2}{\Delta s} & \langle x'^2 \rangle + 2 \frac{\sigma^2}{\Delta s^2} \end{bmatrix},
\end{equation}
where $\sigma$ is the BPM resolution. The true covariances of the jitter can be expressed in terms of its waist parameters as $\langle x^2 \rangle = \epsilon(\beta_w+s_w^2/\beta_w)$, $\langle xx' \rangle = -\epsilon s_w/\beta_w$, and $\langle x'^2 \rangle = \epsilon/\beta_w$, where $\epsilon$ is the geometric jitter emittance and $s_w$ is the distance from the upstream BPM to the jitter waist. 

The measured jitter emittance for this finite BPM resolution is given by the determinant of Eq.~\ref{eq:CovarianceMatrix}
\begin{equation}
    \label{eq:MeasuredEmittance}
    \hat{\epsilon}^2 = \epsilon^2 + \frac{\sigma^2}{ \Delta s^2}\frac{\epsilon}{\beta_w}\left( s_w^2 + (\Delta s - s_w)^2 + 2\beta_w^2 \right) + \frac{\sigma^4}{\Delta s^2}.
\end{equation}
Employing the same logic as in Sec.~\ref{sec:PhaseSpaces} (for Eqs.~\ref{eq:MismatchedWaistBeta} and \ref{eq:MismatchedWaistLocation}), we can find the measured waist beta function from the inverse of the measured gamma function
\begin{equation}
    \label{eq:MeasuredWaistBetaFunction}
    \hat{\beta}_w = \frac{\hat{\epsilon}\beta_w}{\epsilon+2\frac{\beta_w\sigma^2}{\Delta s^2}},
\end{equation}
where $\hat{\epsilon}$ can be substituted from Eq.~\ref{eq:MeasuredEmittance}, as well as the measured waist location from the ratio of the measured alpha and gamma functions
\begin{equation}
    \label{eq:MeasuredWaistLocation}
    \hat{s}_w = \frac{s_w + \frac{\beta_w\sigma^2}{\epsilon\Delta s}}{1 + 2\frac{\beta_w\sigma^2}{\epsilon\Delta s^2}}.
\end{equation}

\begin{figure}[t]
	\centering\includegraphics[width=0.98\linewidth]{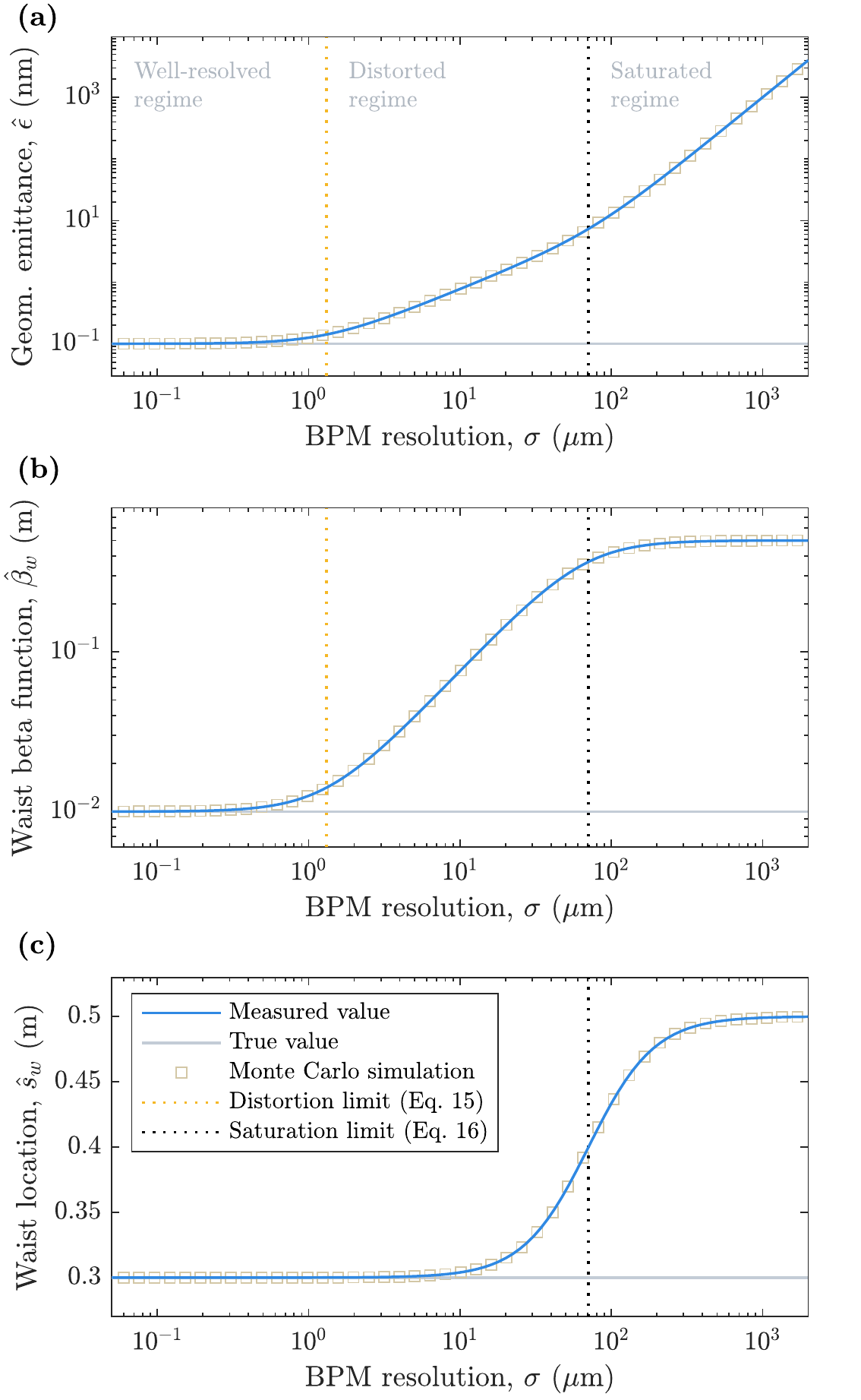}
	\caption{Simulated two-BPM measurements over a large range of finite BPM resolutions. The BPMs are spaced 1~m apart and the jitter is focused 0.3~m from the upstream BPM with a 10~mm beta function and 0.1~mm mrad normalized emittance. Monte Carlo simulations (average of 10$^7$ shots per resolution) demonstrate that this analytic model is exact.}
    \label{fig:Fig3}
\end{figure}

Equations~\ref{eq:MeasuredEmittance}--\ref{eq:MeasuredWaistLocation} establish three resolution regimes: (1)~well-resolved, (2)~distorted, and (3)~fully saturated. To avoid any distortion whatsoever, the BPM resolution must be better than
\begin{equation}
    \label{eq:DistortionLimit}
    \sigma \ll \sqrt{\frac{\epsilon \beta_w \Delta s^2}{s_w^2 + (\Delta s - s_w)^2 + 2\beta_w^2}},
\end{equation}
found by requiring the quadratic $\sigma^2$-term in Eq.~\ref{eq:MeasuredEmittance} to be smaller than the constant $\epsilon^2$-term. To avoid saturation (i.e., noise dominating the signal), the resolution should be better than
\begin{equation}
    \label{eq:SaturationLimit}
    \sigma \ll \sqrt{\frac{\epsilon \Delta s^2}{2\beta_w}},
\end{equation}
found by demanding the $\sigma^2$-term in the denominator of Eq.~\ref{eq:MeasuredWaistBetaFunction} be smaller than the $\epsilon$-term. Encouragingly, the measurement of the waist location is not affected by the distortion limit, and instead only by the significantly larger saturation limit. This is because the waist location is only related to the phase-space correlation and not its area. These regimes are demonstrated by the example in Fig.~\ref{fig:Fig3}, which also shows exact agreement with Monte Carlo simulations of two finite-resolution BPMs.

In a typical case where the waist beta function is small compared to the BPM separation ($\beta_w \ll \Delta s$) and the waist is approximately half way between the BPMs ($s_w \approx \Delta s/2$), Eq.~\ref{eq:DistortionLimit} simplifies to $\sigma \ll \sqrt{2\epsilon\beta_w}$---therefore the BPM resolution should be smaller than the position jitter at the waist. This limit informs the choice of BPM technology required for the application in question.

\subsection{Overcoming the resolution limit}
For matching into a plasma accelerator with mm-scale beta functions and sub-{\textmu}m jitter emittances, a very high BPM resolution is required. Ideally, this is achieved using state-of-the-art cavity BPMs, which can provide sub-100~nm resolution (depending on the charge distribution) \cite{KimPRAB2012,BettIPAC2018}. However, if such BPMs are not available, measuring the waist beta function may require going beyond the resolution limit. This is in principle possible to do, if the emittance of the jitter is already known.

Just like the emittance of the beam, the jitter emittance is preserved in a linear-optics lattice (assuming it contains no significant jitter sources). Therefore one of two alternative measurement techniques can be utilized: (1) Relax the strength of the focusing until the jitter waist is well resolved---giving different Twiss parameters, but the same emittance. (2) Simultaneously perform a similar measurement with two other BPMs just upstream or just downstream, where the focusing is relaxed compared to the focus region. The first method requires the jitter emittance to persist in time, whereas the second requires it to persist in space. Both methods assume negligible chromaticity or that energy slices are measured separately (see Sec.~\ref{sec:SliceMeasurements}).

When the jitter emittance is known, the analysis simplifies greatly. The waist beta function can be calculated using
\begin{equation}
    \beta_w = \frac{\epsilon \Delta s^2}{\langle (x_2-x_1)^2 \rangle},
\end{equation}
based on the variance of the angle jitter (Eq.~\ref{eq:BPMangle}), and the waist location is simply
\begin{equation}
    \label{eq:PhaseSpaceCorrelationWaist}
    s_w = \frac{\Delta s}{1 - \frac{\partial x_2}{\partial x_1}},
\end{equation}
where $\frac{\partial x_2}{\partial x_1}$ is the slope of the correlation between the two BPM readings.

\section{Measurements at FLASHForward}

Experimental demonstration of the two-BPM method was performed at the FLASHForward facility at DESY, which uses a 1~GeV electron beam from the FLASH free-electron-laser facility \cite{AckermannNatPhot2007}. FLASH provides high-charge (up to 1~nC), low-emittance (1~mm~mrad) bunches with relatively small centroid jitter. After an approximately 150~m long linac, the bunches are diverted into the FLASHForward beamline. Here, a dispersive section allows for advanced energetic collimation \cite{SchroederIOP2020}, then a final-focusing section [depicted in Fig.~\ref{fig:Fig1}(b)] tightly focuses the beam into a plasma accelerator. Downstream of the plasma is a suite of beam diagnostics, in particular a dipole spectrometer with quadrupoles for point-to-point beam imaging.

\subsection{Comparison to quadrupole scans}

\begin{figure}[t]
	\centering\includegraphics[width=0.98\linewidth]{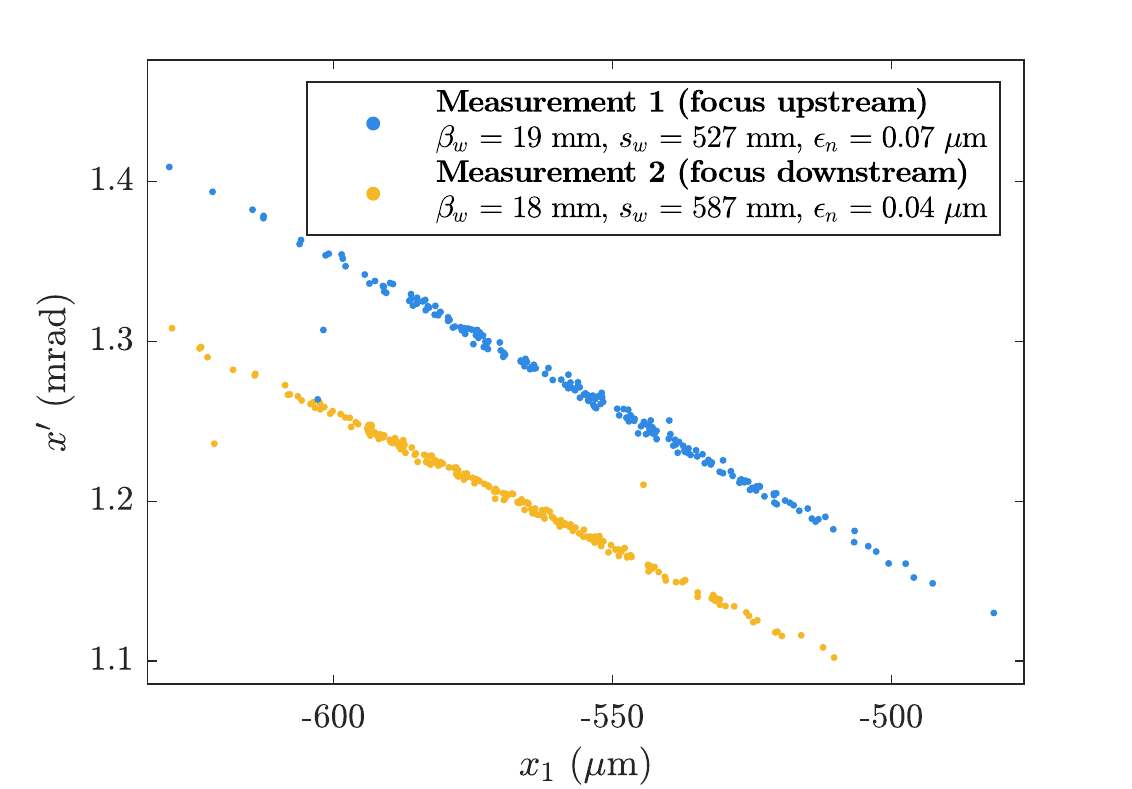}
	\caption{Measured jitter phase spaces at the location of the upstream BPM for two different optics settings, indicating cm-scale waist beta functions focused at two waist locations 60~mm apart. A small distortion from a finite BPM resolution was taken into account when calculating the jitter parameters. These measurements should be compared to the corresponding quadrupole scans in Fig.~\ref{fig:Fig5}. Each dataset consists of 210 shots, giving an estimated relative error of 7\%.}
    \label{fig:Fig4}
\end{figure}

\begin{figure*}[t]
	\centering\includegraphics[width=0.98\textwidth]{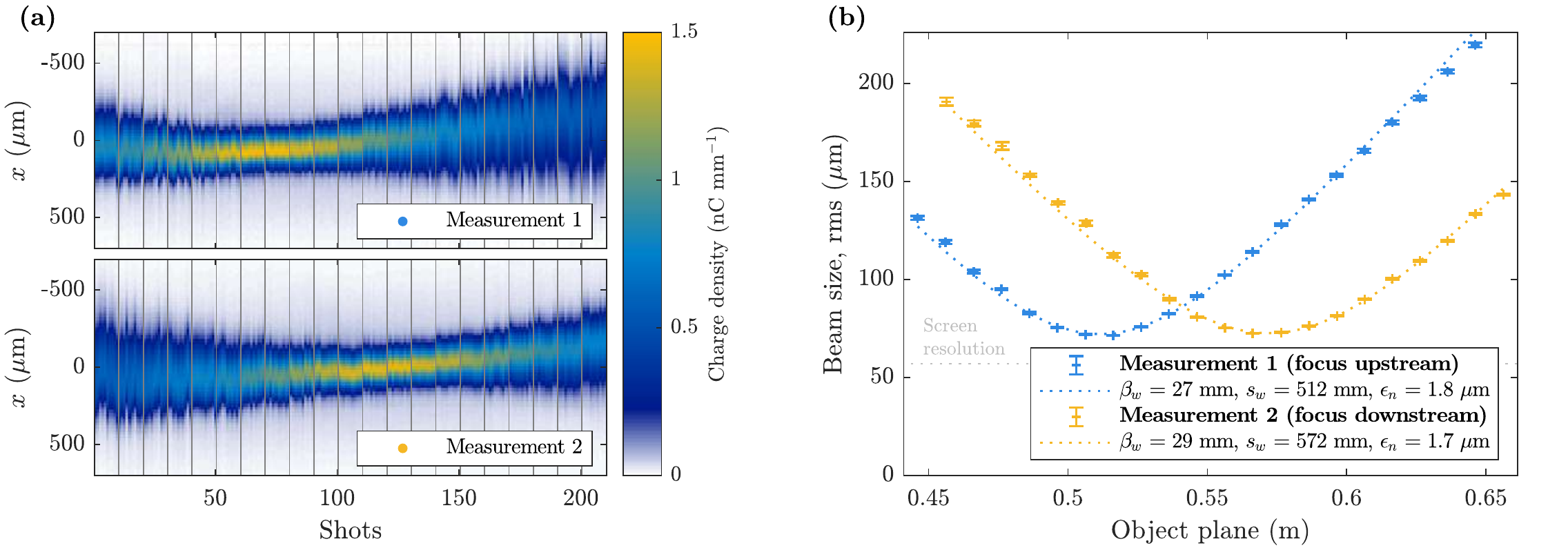}
	\caption{Quadrupole scans on a downstream spectrometer screen performed during the two-BPM measurements in Fig.~\ref{fig:Fig4}, (a) imaging the beam from a range of object planes around the beam waist. (b) The variation of horizontally projected beam size for each object plane indicates that the beam was focused to a small waist beta function (27--29~mm) close to the center of the focus region---only moderately mismatched from measured jitter. The screen resolution was accounted for in the calculation of beam parameters. Error bars represent the standard error of the mean.}
    \label{fig:Fig5}
\end{figure*}

To test the assumptions in Sec.~\ref{sec:PhaseSpaces} and the applicability of the method, a detailed comparison of the measured jitter and beam phase spaces was performed. A strong-focusing optic was set up to focus bunches with an energy of 678~MeV and charge 290~pC down to a cm-scale beta function at the location of the plasma accelerator module (which had been removed from the beam path). Surrounding this focus region were two cavity BPMs \cite{LipkaIBIC2015,NolleIBIC2019} with a resolution of 0.9~{\textmu}m, separated by 1.073~m, and approximately equidistant from the nominal focus point. 

Two datasets were collected, using slightly different final-focusing optics with the beam focused at two locations 60~mm apart. Figure~\ref{fig:Fig4} shows the measured jitter phase space for each of these two settings using the two-BPM method. The presence of outliers (as seen in Fig.~\ref{fig:Fig4}) can significantly skew the calculation of phase-space parameters, and thus an outlier-cleaning method was applied: (i) translate the jitter to the waist location from the BPM correlation (Eq.~\ref{eq:PhaseSpaceCorrelationWaist}), (ii) perform Gaussian fits of both the $x$ and $x'$ distributions, (iii) remove all shots beyond $\pm5\sigma$, and then (iv) undo the translation from (i). Finally, a small distortive effect from the finite BPM resolution was removed by numerically solving Eqs.~\ref{eq:MeasuredEmittance}--\ref{eq:MeasuredWaistLocation} for the true jitter-phase-space parameters.

At the same time, an object-plane scan was performed with the downstream quadrupoles (after the second BPM), imaging the beam onto a LANEX screen with a resolution of 57~{\textmu}m. Figure~\ref{fig:Fig5} shows the corresponding measurement of the beam waist. Note that the spectrometer limits the measurement to the horizontal plane, as the dipole disperses vertically. No chromaticity was observed on the screen.

The two-BPM measurement agrees with the quadrupole scan measurement to an acceptable level. The waist beta function of the jitter (18--19~mm) differs from that of the beam (27--29~mm) by about 35\%, and the jitter waist location is offset from the beam waist location by \mbox{15~mm}---on the same scale as the waist beta function, as expected. Based on these numbers, the beam--jitter mismatch parameter was calculated to be $\mathcal{M} = 2.1$--$2.2$, implying that the phase space of the jitter was indeed closely matched to that of the beam. If used to match into a plasma accelerator (where the jitter phase space would be matched), the expected emittance growth of the beam from mismatching (Eq.~\ref{eq:MismatchEmitGrowth}) would be 28--33\%.

\begin{figure}[t]
	\centering\includegraphics[width=0.98\linewidth]{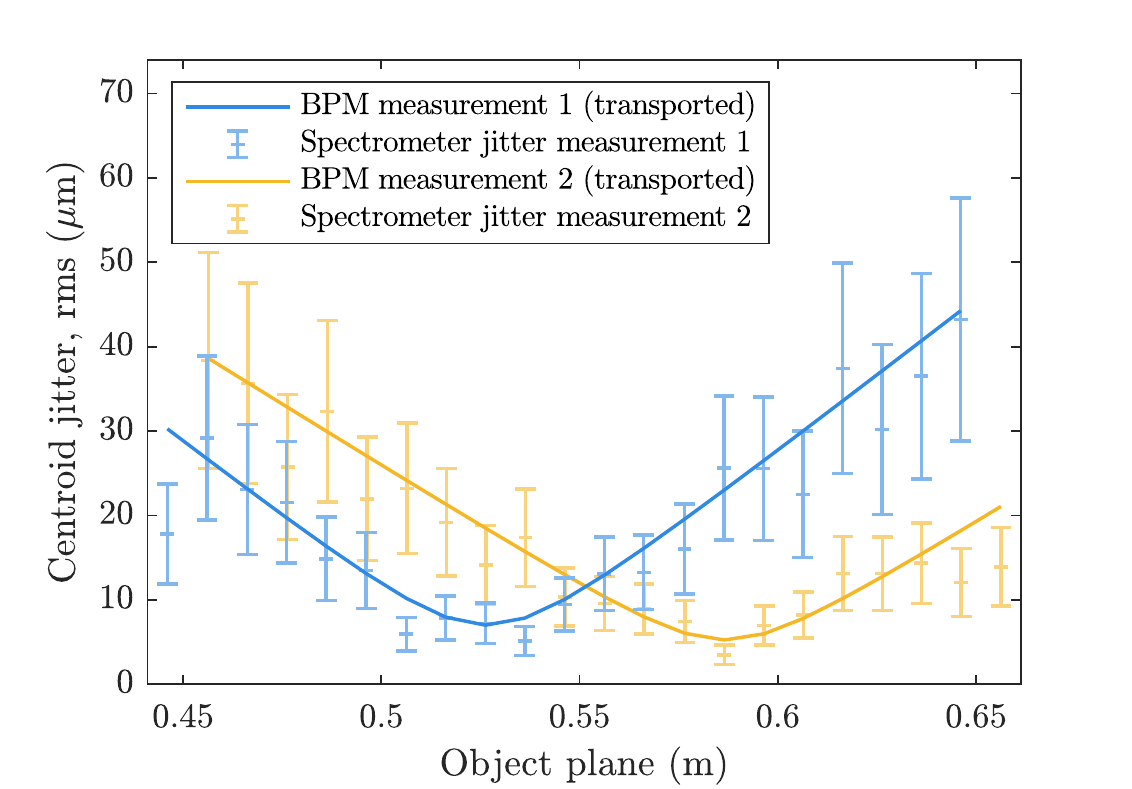}
	\caption{Comparison of the centroid jitter measured on the spectrometer screen and with the two-BPM method (artificially transported through the quadrupoles to the spectrometer location). Error bars represent the standard error of the mean. A statistically significant agreement is observed, verifying the accuracy of both measurement methods.}
    \label{fig:Fig6}
\end{figure}

As an additional cross-check of the jitter measurements, the centroid jitter was also measured directly on the spectrometer (see Fig.~\ref{fig:Fig6}). This was used to verify the accuracy of the distances and quadrupole field strength calibrations used for the quadrupole scans, as well as to fine-tune the value of the BPM resolution and calibrations.

\subsection{Slice-by-slice measurements}
\label{sec:SliceMeasurements}

Chromaticity, where Twiss parameters change with energy \cite{ZyngierLAL1977,MontagueLEP1979}, can be a concern when tightly focusing beams of finite energy spread \cite{LindstromPRAB2016}. This is especially important in energy-chirp-based two-bunch experiments where a trailing bunch needs to be exactly matched into the plasma wake behind a different-energy driver bunch. 

\begin{figure}[t]
	\centering\includegraphics[width=0.98\linewidth]{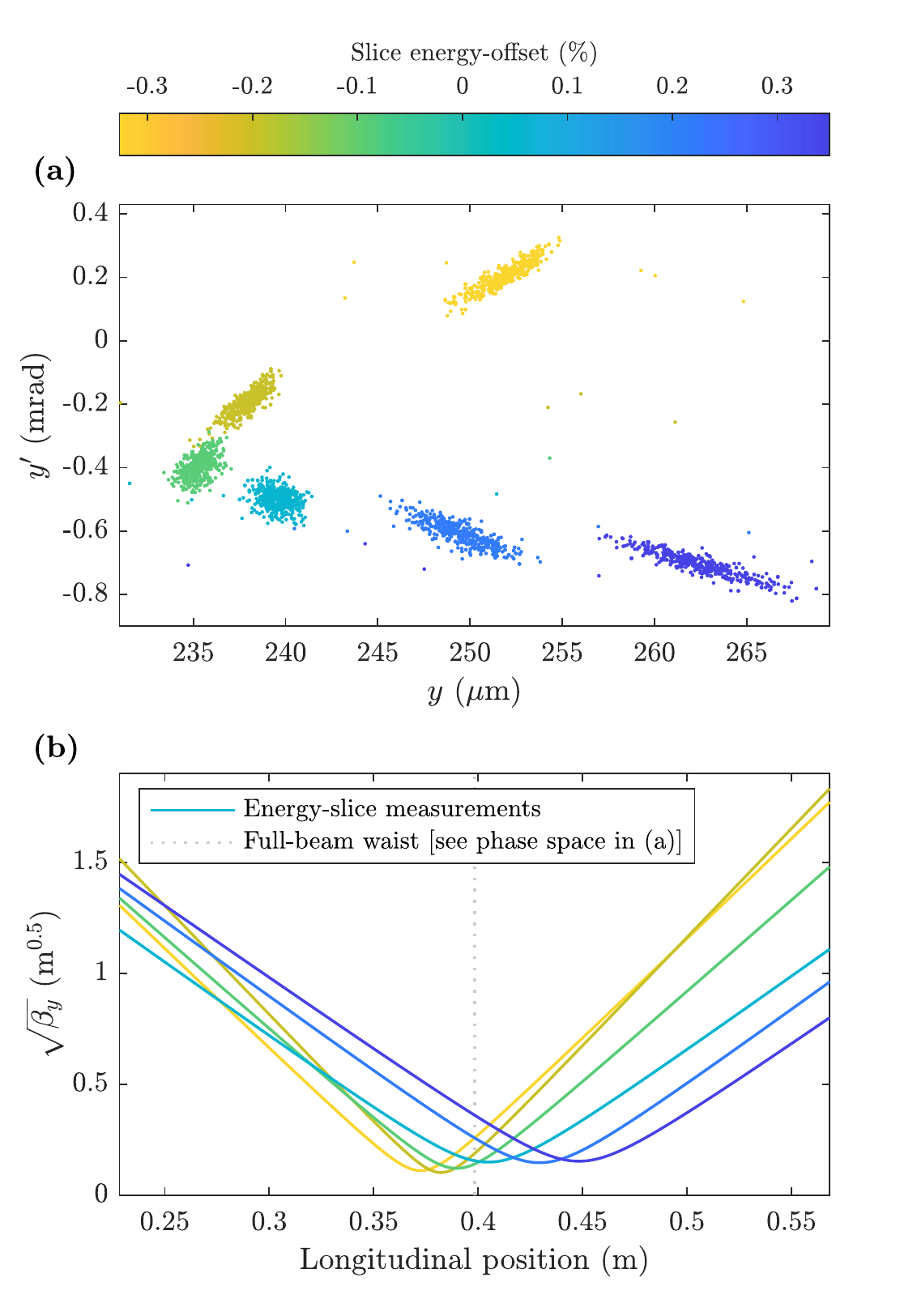}
	\caption{Chromaticity measurement using an energy-slice scan. (a) Jitter phase space for a range of energy slices. Significant dispersion (energy-dependent offsets) can also be observed. Each step consists of 500 shots. (b) The corresponding evolution of the beta function for each slice in the focus region, indicating a highly chromatic focus.}
    \label{fig:Fig7}
\end{figure}

Measuring chromaticity with the two-BPM technique requires it to be combined with an energy filter---each energy slice sufficiently narrow to have an achromatic focus. At FLASHForward this is accomplished using an energetic collimator \cite{SchroederIOP2020}. Moving both the high- and low-energy collimators together, thin slices with 0.1\% root-mean-square (rms) energy spread could be made. Figure~\ref{fig:Fig7} shows the result of such an energy-slice scan around a mean energy of 1120~MeV, indicating a highly chromatic focus in the vertical plane. The waist beta functions are relatively consistent (10--20~mm) across all slices, whereas the waist location shifted significantly between the highest and lowest energy slice (by 80~mm). In the horizontal plane (not shown in Fig.~\ref{fig:Fig7}), the waist location spanned only 10~mm---considerably less chromatic. This asymmetric chromaticity is expected in a quadrupole-based final-focus system, where the beam is more strongly defocused in one plane (typically the vertical plane) before being focused to a waist.

Taking into account all the information gathered in an energy-slice scan, we can extract a partial 5D beam tomography. As seen in Fig.~\ref{fig:Fig7}(a), the average position and angle of each individual energy slice is also measured, and therefore both the beam centroid and the (emittance-normalized) beam size of each slice is known in both planes. This in-situ tomography allows not only slice-specific matching, but also measurement and removal of any bunch dispersion. For a linearized longitudinal phase space, dispersion corresponds to a bunch tilt or curvature, which in a plasma wake leads to emittance growth \cite{AssmannNIMA1998,LindstromIPAC2016} and potentially a hosing instability \cite{WhittumPRL1991, HuangPRL2007, MehrlingPRL2017}. Finally, combining such a two-BPM tomography with longitudinal-phase-space data from a transverse-deflecting cavity allows a 6D phase space to be reconstructed---important for realistic simulations and detailed optimization of the external injection process.

\section{Conclusions}
We have shown that centroid jitter measured by two BPMs can be used to quickly estimate Twiss parameters in a region of strong focusing. While being an approximate measurement, it can significantly speed up the complex and delicate beam setup procedure needed to properly match into a plasma accelerator, and allows noninvasive online monitoring of the beam focus. Experiments were successfully performed at FLASHForward to verify this technique, by comparing the two-BPM measurement to a conventional quadrupole scan. Already in routine use for plasma-wakefield experiments at FLASHForward, it is clear that the power of this method lies in its simplicity.

\begin{acknowledgments}
    The authors would like to thank K.~Fl{\"o}ttmann, S.~Karstensen, K.~Ludwig, F.~Marutzky, A.~Rahali, A.~Schleiermacher and S.~Thiele, as well as the FLASH management for their scientific and technical support. This work was supported by the Helmholtz ARD program.
\end{acknowledgments}



\begin{thebibliography}{40}

    \bibitem{TajimaDawsonPRL1979}
    T. Tajima and J. M. Dawson,
    Laser electron accelerator,
    \href{https://doi.org/10.1103/PhysRevLett.43.267}
    {Phys. Rev. Lett. \textbf{43}, 267 (1979)}.
    
    \bibitem{ChenPRL1985}
    P. Chen, J. M. Dawson, R. W. Huff and T. Katsouleas, 
    Acceleration of electrons by the interaction of a bunched electron beam with a plasma,
    \href{https://doi.org/10.1103/PhysRevLett.54.693}
    {Phys. Rev. Lett. \textbf{54}, 693 (1985)}.
    
    \bibitem{RuthPA1985}
    R. D. Ruth, A. W. Chao, P. L. Morton and P. W. Wilson,
    A plasma wake field accelerator,
    \href{http://cds.cern.ch/record/157249/files/p171.pdf}
    {Part. Accel. \textbf{17}, 171 (1985)}.
    
    \bibitem{BlumenfeldNature2007}
    I. Blumenfeld, C. E. Clayton, F.-J. Decker, M. J. Hogan, C. Huang, R. Ischebeck \textit{et al.},
    Energy doubling of 42 GeV electrons in a metre-scale plasma wakefield accelerator,
    \href{https://doi.org/10.1038/nature05538}
    {Nature (London) \textbf{445}, 741 (2007)}.
    
    \bibitem{LitosNature2014}
    M. Litos, E. Adli, W. An, C. I. Clarke, C. E. Clayton, S. Corde \textit{et al.},
    High-efficiency acceleration of an electron beam in a plasma wakefield accelerator,
    \href{https://doi.org/10.1038/nature13882}
    {Nature (London) \textbf{515}, 92 (2014)}.
    
    \bibitem{JoshiPT2003}
    C. Joshi and T. Katsouleas,
    Plasma accelerators at the energy frontier and on tabletops,
    \href{https://doi.org/10.1063/1.1595054}
    {Phys. Today \textbf{56}, 6, 47 (2003)}.
    
    \bibitem{LeemansPT2009}
    W. P. Leemans and E. Esarey,
    Laser-driven plasma-wave electron accelerators,
    \href{https://doi.org/10.1063/1.3099645}
    {Phys. Today \textbf{62}, No. 3, 44 (2009)}.
    
    \bibitem{MadeyJAP1971}
    J. M. J. Madey,
    Stimulated emission of bremsstrahlung in a periodic magnetic field,
    \href{https://doi.org/10.1063/1.1660466}
    {J. Appl. Phys. \textbf{42}, 1906 (1971)}.
    
    \bibitem{CouprieJPB2014}
    M. E. Couprie, A. Loulergue, M. Labat, R. Lehe and V. Malka,
    Towards a free electron laser based on laser plasma accelerators,
    \href{https://doi.org/10.1088/0953-4075/47/23/234001}
    {J. Phys. B \textbf{47}, 234001 (2014)}.
    
    \bibitem{RosenzweigNIMA1998}
    J. Rosenzweig, N. Barov, A. Murokh, E. Colby and P. Colestock,
    Towards a plasma wake-field acceleration-based linear collider,
    \href{https://doi.org/10.1016/S0168-9002(98)00186-7}
    {Nucl. Instrum. Methods Phys. Res. A \textbf{410}, 532 (1998)}.
    
    \bibitem{SchroederPhysRevSTAB2010}
    C. B. Schroeder, E. Esarey, C. G. R. Geddes, C. Benedetti and W. P. Leemans,
    Physics considerations for laser-plasma linear colliders,
    \href{https://doi.org/10.1103/PhysRevSTAB.13.101301}
    {Phys. Rev. ST Accel. Beams \textbf{13}, 101301 (2010)}.
    
    \bibitem{AdliSnowmass2013}
    E. Adli, J.-P. Delahaye, S. J. Gessner, M. J. Hogan, T. O. Raubenheimer, W. An, C. Joshi and W. B. Mori,
    A beam driven plasma-wakefield linear collider: from Higgs factory to multi-TeV, SLAC Report No. SLAC-PUB-15426, 2013.
    
    \bibitem{LindstromThesis2019}
    C. A. Lindstr{\o}m,
    Emittance growth and preservation in a plasma-based linear collider,
    \href{http://urn.nb.no/URN:NBN:no-69347}
    {Ph.D. thesis, University of Oslo, 2019}.
    
	\bibitem{SteinkeNature2016}
    S. Steinke, J. van tilborg, C. Benedetti, C. G. R. Geddes, C. B. Schroeder, J. Daniels \textit{et al.},
    Multistage coupling of independent laser-plasma accelerators,
    \href{https://doi.org/10.1038/nature16525}
    {Nature (London) \textbf{530}, 190 (2016)}.
    
    \bibitem{LindstromNIMA2016}
    C. A. Lindstr{\o}m, E. Adli, J. M. Allen, J. P. Delahaye, M. J. Hogan, C. Joshi, P. Muggli, T. O. Raubenheimer and V. Yakimenko, 
    Staging optics considerations for a plasma wakefield acceleration linear collider,
    \href{https://doi.org/10.1016/j.nima.2015.12.065}
    {Nucl. Instrum. Methods Phys. Res. A \textbf{829}, 224 (2016)}.

    \bibitem{MehrlingPRAB2012}
    T. Mehrling, J. Grebenyuk, F. S. Tsung, K. Floettmann and J. Osterhoff,
    Transverse emittance growth in staged laser-wakefield acceleration,
    \href{https://doi.org/10.1103/PhysRevSTAB.15.111303}
    {Phys. Rev. ST Accel. Beams \textbf{15}, 111303 (2012)}.
    
    \bibitem{CourantSnyderAoP1957}
    E. D. Courant and H. S. Snyder,
    Theory of the alternating-gradient synchrotron,
    \href{https://doi.org/10.1016/0003-4916(58)90012-5}
    {Ann. Phys. \textbf{3}, 1 (1958)}.

    \bibitem{MarshPAC2005}
    K. A. Marsh, C. E. Clayton, D. K. Johnson, C. Huang, C. Joshi, W. Lu \textit{et al.},
    Beam matching to a plasma wake field accelerator using a ramped density profile at the plasma boundary,
    \href{https://doi.org/10.1109/PAC.2005.1591234}
    {\textit{Proceedings of PAC2005, Knoxville, TN, USA} (JACoW, Geneva, 2005), p. 2702}.
    
    \bibitem{DornmairPRAB2015}
    I. Dornmair, K. Floettmann and A. R. Maier,
    Emittance conservation by tailored focusing profiles in a plasma accelerator,
    \href{https://doi.org/10.1103/PhysRevSTAB.18.041302}
    {Phys. Rev. ST Accel. Beams \textbf{18}, 041302 (2015)}.
    
    \bibitem{ZhaoPRAB2020}
    Y. Zhao, W. An, X. Xu, F. Li, L. Hildebrand, M. J. Hogan, V. Yakimenko, C. Joshi and W. B. Mori,
    Emittance preservation through density ramp matching sections in a plasma wakefield accelerator,
    \href{https://doi.org/10.1103/PhysRevAccelBeams.23.011302}
    {Phys. Rev. Accel. Beams \textbf{23}, 011302 (2020)}.
    
    \bibitem{AschikhinNIMA2016}
    A. Aschikhin, C. Behrens, S. Bohlen, J. Dale, N. Delbos, L. di Lucchio \textit{et al.},
    The FLASHForward facility at DESY,
    \href{https://doi.org/10.1016/j.nima.2015.10.005}
    {Nucl. Instrum. Methods Phys. Res. A \textbf{806}, 175 (2016)}.
    
    \bibitem{DArcyRSTA2019}
    R. D'Arcy, A. Aschikhin, S. Bohlen, G. Boyle, T. Br{\"u}mmer, J. Chappell \textit{et al.},
    FLASHForward: plasma wakefield accelerator science for high-average-power applications,
    \href{https://doi.org/10.1098/rsta.2018.0392}
    {Philos. Trans. Royal Soc. A \textbf{377}, 20180392 (2019)}.
    
    \bibitem{BalikIPAC2017}
    G. Balik, B. Aimard, L. Brunetti and B. Caron,
    Proof of concept of CLIC final focus quadrupoles stabilization,
    \href{https://doi.org/10.18429/JACoW-IPAC2017-TUPAB001}
    {Proceedings of IPAC2017, Copenhagen, Denmark (JACoW, Geneva, 2017), p. 1290}.
    
    \bibitem{BettNIMA2018}
    D. R. Bett, C. Charrondi{\`e}re, M. Pateckia, J. Pfingstner, D. Schulte, R. Tom{\'a}s \textit{et al.},
    Compensation of orbit distortion due to quadrupole motion using feed-forward control at KEK ATF,
    \href{https://doi.org/10.1016/j.nima.2018.03.037}
    {Nucl. Instrum. Methods Phys. Res. A \textbf{895}, 10 (2018)}.
    
    \bibitem{SandsSLAC1991}
    M. Sands,
    A beta mismatch parameter,
    SLAC Report No. SLAC-AP-85, 1991.
    
    \bibitem{YaminIPAC2019}
    S. Yamin, R. W. Assmann, U. Dorda, F. Lemery, B. Marchetti, E. Panofski and P. A. Walker,
    Design considerations for permanent magnetic quadrupole triplet for matching into laser driver wake field acceleration experiment at SINBAD,
    \href{https://doi.org/10.18429/JACoW-IPAC2019-MOPGW027}
    {Proceedings of IPAC2019, Melbourne, Australia (JACoW, Geneva, 2019), p. 143}.
    
    \bibitem{KimPRAB2012}
    Y. I. Kim, R. Ainsworth, A. Aryshev, S. T. Boogert, G. Boorman, J. Frisch \textit{et al.},
    Cavity beam position monitor system for the Accelerator Test Facility 2,
    \href{https://doi.org/10.1103/PhysRevSTAB.15.042801}
    {Phys. Rev. ST Accel. Beams \textbf{15}, 042801 (2012)}.
    
    \bibitem{BettIPAC2018}
    D. R. Bett, N. Blaskovic Kraljevic, R. M. Bodenstein, T. Bromwich, P. N. Burrows, G. B. Christian \textit{et al.},
    Performance of nanometre-level resolution cavity beam position monitors at ATF2,
    \href{https://doi.org/10.18429/JACoW-IPAC2018-TUZGBD5}
    {Proceedings of IPAC2018, Vancouver, BC, Canada (JACoW, Geneva, 2018), p. 1212}.

    \bibitem{AckermannNatPhot2007}
    W. Ackermann, G. Asova, V. Ayvazyan, A. Azima, N. Baboi, J. B{\"a}hr \textit{et al.},
    Operation of a free-electron laser from the extreme ultraviolet to the water window,
    \href{https://doi.org/10.1038/nphoton.2007.76}
    {Nat. Photonics \textbf{1}, 336 (2007)}.
    
    \bibitem{SchroederIOP2020}
    S. Schr{\"o}der, K. Ludwig, A. Aschikhin, R. D'Arcy, M. Dinter, P. Gonzalez \textit{et al.},
    Tunable and precise two-bunch generation at FLASHForward,
    \href{https://arxiv.org/abs/2005.12071}{arXiv:2005.12071} 
    [J. Phys.: Conf. Ser. (to be published)].
    
    \bibitem{LipkaIBIC2015}
    D. Lipka, N. Baboi, D. Noelle, G. Petrosyan, S. Vilcins, R. Baldinger \textit{et al.},
    FLASH undulator BPM commissioning and beam characterization results,
    \href{http://jacow.org/IBIC2014/papers/tupf07.pdf}
    {\textit{Proceedings of IBIC2014, Monterey, CA, USA} (JACoW, Geneva, 2015), p. 315}.
    
    \bibitem{NolleIBIC2019}
    D. N{\"o}lle \textit{et al.},
    The diagnostic system at the European XFEL; commissioning and first user operation,
    \href{https://doi.org/10.18429/JACoW-IBIC2018-TUOA01}
    {\textit{Proceedings of IBIC2018, Shanghai, China} (JACoW, Geneva, 2019), p. 162}.
    
    \bibitem{ZyngierLAL1977}
    H. Zyngier,
    Strategy for correcting for chromaticity,
    \href{https://lib-extopc.kek.jp/preprints/PDF/1978/7801/7801139.pdf}
    {\textit{LAL-77/35} (Laboratoire de l'Acc{\'e}l{\'e}rateur Lin{\'e}aire, Orsay, 1977)}.
    
     \bibitem{MontagueLEP1979}
    B. W. Montague,
    Linear optics for improved chromaticity correction,
    \href{http://cds.cern.ch/record/67243}
    {\textit{LEP Note 165} (CERN, Geneva, 1979)}.
    
    \bibitem{LindstromPRAB2016}
    C. A. Lindstr{\o}m and E. Adli, 
    Design of general apochromatic drift-quadrupole beam lines,
    \href{https://doi.org/10.1103/PhysRevAccelBeams.19.071002}
    {Phys. Rev. Accel. Beams \textbf{19}, 071002 (2016)}.
    
    \bibitem{AssmannNIMA1998}
    R. Assmann and K. Yokoya,
    Transverse beam dynamics in plasma-based linacs,
    \href{https://doi.org/10.1016/S0168-9002(98)00187-9}
    {Nucl. Instrum. Methods Phys. Res. A \textbf{410}, 544 (1998)}.
    
    \bibitem{LindstromIPAC2016}
    C. A. Lindstr{\o}m, E. Adli, J. Pfingstner, E. Mar{\'i}n and D. Schulte,
    Transverse tolerances of a multi-stage plasma wakefield accelerator,
    \href{https://doi.org/10.18429/JACoW-IPAC2016-WEPMY009}
    {\textit{Proceedings of IPAC2016, Busan, Korea} (JACoW, Geneva, 2016), p. 2561}.
    
     \bibitem{WhittumPRL1991}
    D. H. Whittum, W. M. Sharp, S. S. Yu, M. Lampe and G. Joyce,
    Electron-hose instability in the ion-focused regime,
    \href{https://doi.org/10.1103/PhysRevLett.67.991}
    {Phys. Rev. Lett. \textbf{67}, 991 (1991)}.
    
    \bibitem{HuangPRL2007}
    C. Huang, W. Lu, M. Zhou, C. E. Clayton, C. Joshi, W. B. Mori \textit{et al.},
    Hosing instability in the blow-out regime for plasma-wakefield acceleration,
    \href{https://doi.org/10.1103/PhysRevLett.99.255001}
    {Phys. Rev. Lett. \textbf{99}, 255001 (2007)}.
    
    \bibitem{MehrlingPRL2017}
    T. J. Mehrling, R. A. Fonseca, A. Martinez de la Ossa and J. Vieira,
    Mitigation of the hose instability in plasma-wakefield accelerators,
    \href{https://doi.org/10.1103/PhysRevLett.118.174801}
    {Phys. Rev. Lett. \textbf{118}, 174801 (2017)}.
    
\end{thebibliography}
\end{document}